\documentclass[traditabstract]{aa} 

\usepackage{pgf}
\usepackage{graphicx}
\usepackage{txfonts}
\usepackage{natbib}
\usepackage{epsf}
\usepackage{rotating}
\usepackage{graphics}
\usepackage{verbatim}
\usepackage{longtable}

\def \sophie{SOPHIE}
\def \kepler{\emph{Kepler}}

\def\kepler{\emph{Kepler}}
\def \MJ{M$_{\mathrm{Jup}}$}
\def \RJ{R$_{\mathrm{Jup}}$}
\def \Msun{M$_{\odot}$}
\def \Rsun{R$_{\odot}$}

\def \kms{km\,s$^{-1}$}
\def \ms{m\,s$^{-1}$}
\def \1s{$1\,\sigma$}

\def \t0{T$_0$}

\def \189{KOI-189}
\def \686{KOI-686}

\begin{document}

\title{SOPHIE velocimetry of \kepler\ transit candidates \thanks{Based on observations collected with the {\it SOPHIE}  spectrograph on the 1.93-m telescope at Observatoire de Haute-Provence (CNRS), France (programs 11A.PNP.MOUT and 11B.PNP.MOUT). Tables~\ref{table.rv189},~\ref{table.rv686}, and \ref{table.priors} are only available online.}}
            
\subtitle{XIII. \189~B and \686~B: two very low-mass stars in long-period orbits.}

\author{
R.~F.~D\'iaz\inst{1,2}, G.~Montagnier\inst{3,4}, J. Leconte\inst{5}, A.~S.~Bonomo\inst{6}, M.~Deleuil\inst{2}, J.~M.~Almenara\inst{2}, S.~C.~C.~Barros\inst{2}, F.~Bouchy\inst{3,4}, G.~Bruno\inst{2}, C.~Damiani\inst{2}, G.~H\'ebrard\inst{3,4}, C.~Moutou\inst{2,7},  A.~Santerne\inst{2,8,9}
}

\institute{
Observatoire Astronomique de l'Universit\'e de Gen\'eve, Chemin des Maillettes 51, 1290 Versoix, Switzerland\and
Aix Marseille Universit\'e, CNRS, LAM (Laboratoire d'Astrophysique de Marseille) UMR 7326, 13388, Marseille, France \and 
Institut d'Astrophysique de Paris, UMR7095 CNRS, Universit\'e Pierre \& Marie Curie, 98bis boulevard Arago, 75014 Paris, France \and 
Observatoire de Haute-Provence, CNRS/OAMP, 04870 Saint-Michel-l'Observatoire, France \and
Canadian Institute for Theoretical Astrophysics, 60 St. George St., University of Toronto, Toronto, ON M5S 3H8, Canada\and
INAF - Osservatorio Astrofisico di Torino, via Osservatorio 20, 10025, Pino Torinese, Italy\and
Canada France Hawaii Telescope Corporation, Kamuela, USA\and
Centro de Astrof\'isica, Universidade do Porto, Rua das Estrelas, P-4150-762 Porto, Portugal\and
Instituto de Astrof\'isica e Ci\^{e}ncias do Espa\c co, Universidade do Porto, CAUP, Rua das Estrelas, PT4150-762 Porto, Portugal
}

 \date{}
      
\abstract{We present the radial-velocity follow-up of two \kepler\ planetary transiting candidates (\189\ and \686) carried out with the SOPHIE spectrograph at the Observatoire de Haute Provence. These data promptly discard these objects as viable planet candidates and show that the transiting objects are in the regime of very low-mass stars, where a strong discrepancy between observations and models persists for the mass and radius parameters. By combining the SOPHIE spectra with the \kepler\ light curve and photometric measurements found in the literature, we obtain a full characterization of the transiting companions, their orbits, and their host stars. The two companions are in significantly eccentric orbits with relatively long periods (30 days and 52.5 days), which makes them suitable objects for a comparison with theoretical models, since the effects invoked to understand the discrepancy with observations are weaker for these orbital distances. \189\ B has a mass $M = 0.0745\pm0.0033$ \Msun\ and a radius $R = 0.1025\pm0.0024$ \Rsun. The density of \189 B is significantly lower than expected from theoretical models for a system of its age. We explore possible explanations for this difference. \189\ B is the smallest hydrogen-burning star with such a precise determination of its fundamental parameters. \686 B is larger and more massive ($M = 0.0915\pm0.0043$ \Msun; $R = 0.1201\pm 0.0033$ \Rsun), and its position in the mass-radius diagram agrees well with theoretical expectations. }

\authorrunning{D\'iaz et al.}
\titlerunning{Two very low-mass stars in long-period orbits.}

\maketitle

\begin{figure*}[t]
\begin{center}
\includegraphics[width=\textwidth]{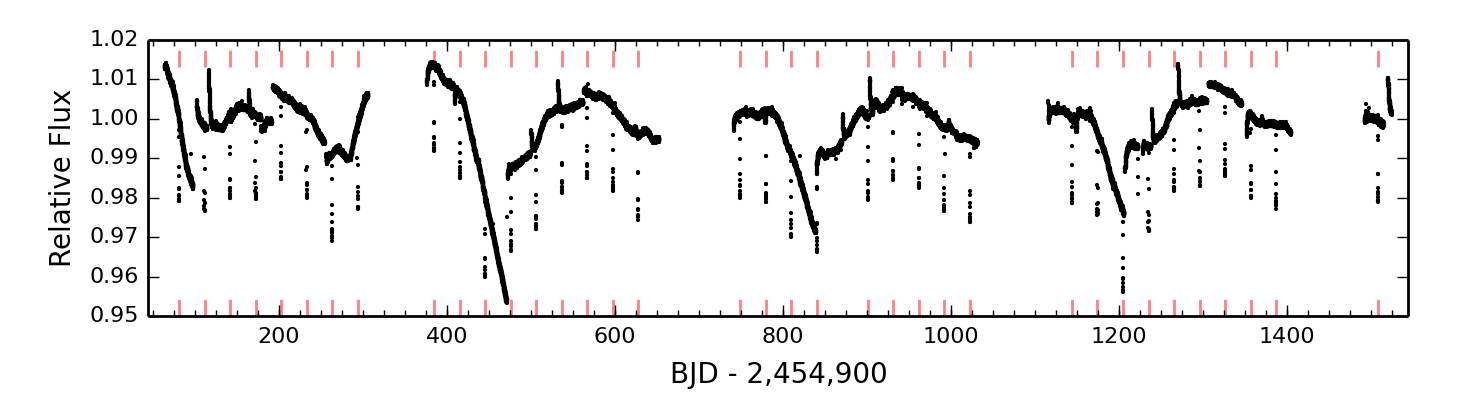}
\includegraphics[width=\textwidth]{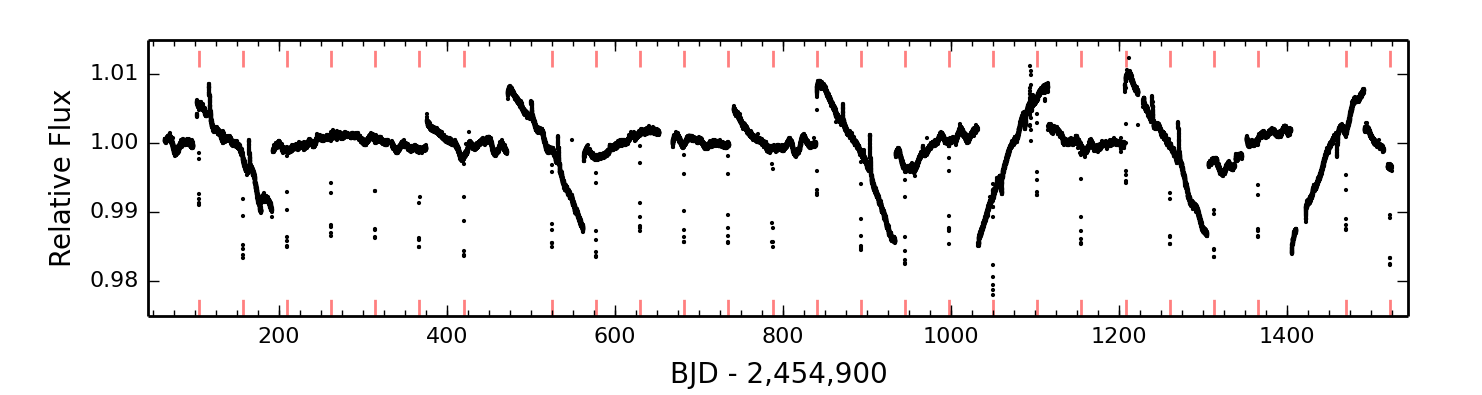}
\caption{Uncorrected \kepler\ light curves for \189 (top) and \686 (bottom). The flux has been scaled to the median value of each quarter. The positions of the transits are shown with vertical red dashed lines. Note the missing parts of the light curve of \189 due to the failure of CCD module 3. \label{fig.rawlcs}}
\end{center}
\end{figure*}

\section{Introduction}
The observed mass-radius relation of low-mass stars is still poorly understood \citep[see][for a recent review]{baraffe2014}. The observed radii of low-mass stars are between 5\% and 10\% larger than predicted by models \citep[e.g.,][]{morales2010}. At present, the inhibition of convection by magnetic fields caused by an increased stellar rotation rate seem the most likely reasons for the discrepancy \citep[e.g.,][]{lopez-morales2007, morales2010, mullanmacdonald2001, chabrier2007}. Indeed, eclipsing binaries tend to be in short-period orbits, and tidal forces tend to synchronize their rotation and orbital periods, which in turn increases their magnetic activity.

The \kepler\ mission is dedicated to finding transiting planets around mainly solar-type stars. Its unprecedented precision and virtually uninterrupted observing has permitted detecting the smallest planets known to date \citep[e.g.,][]{fressin2012, barclay2013} in long-period orbits \citep[e.g.,][]{borucki2012}. \kepler\ photometry is also producing important results in stellar astrophysics. For example, the triply eclipsing triple system KOI-126 \citep{carter2011} provides stringent constrains to models of stellar interiors and tidal interactions. Curiously, the masses and radii of the KOI-126 components agree well with the Dartmouth \citep{dotter2008} stellar tracks, in spite of the short orbital period of the system.

A good understanding of the properties of low-mass stars is fundamental for exoplanetary science, since the planet parameters directly depend on the stellar masses and radii. In particular, very low-mass stars are advantageous targets to search for extrasolar planets \citep[see, e.g.,][]{charbonneau2009}. Understanding their properties should be prioritized.

Here we present the radial velocity follow-up of two \kepler\ planetary candidates whose transits were first reported by \citet{borucki2011} (\189\ and \686). SOPHIE radial-velocity follow-up promptly detected the km/s-amplitude reflex motion of the host stars, which reveals that the transiting objects are actually very low-mass stars in eccentric, long-period (30 and 52 days) orbits. These objects were not included in the determination of the false-positive rate of giant planets of \citet{santerne2012}, since their periods are longer than the cut-off used in that article. 

Combined with the \kepler\ light curves, SOPHIE data permit precise determination of the masses and radii of these objects and their host stars. As discussed in Sect.~\ref{sect.conclusions}, tidal spin-synchronization is expected to be inefficient for these long-period objects, which are expected to be more similar to isolated low-mass stars. These objects are therefore important assets for testing theoretical models. The remainder of the paper is organized as follows: Sect.\ \ref{sect.data} presents the datasets used in the analysis and the data reduction procedure; Sect.\ \ref{sect.starparam} describes the spectroscopic analysis performed to obtain the host stars atmospheric parameters and the attempts to measure their rotation periods from the \kepler\ light curves; Sect.\ \ref{sect.modeling} describes the model used to fit the data, and the Markov chain Monte Carlo algorithm used to sample the posterior distribution of the model parameters; the results are presented in Sect.\ \ref{sect.results} and are summarized and discussed in Sect.\ \ref{sect.conclusions}.

\section{Observations and data reduction \label{sect.data}}
\subsection{\emph{Kepler} light curves}

KIC11391018 and KIC7906882 were observed by \kepler\ in long cadence (LC, i.e., with a sampling of one point every 29.4 minutes) starting in quarter 1, and until the failure of the second reaction wheel in quarter 17. Periodic transits with amplitudes of $\sim2$\% and $\sim1.4$\% and shapes resembling that of planetary transits were promptly detected after the first quarter of observations. After passing all pre-spectroscopic vetting tests \citep[see][]{batalha2010}, the transiting objects earned the title of Kepler Object of Interest (KOI) and were assigned the names \189.01 and \686.01. The host stars are referred to as \189\ and \686. During observing season 2, \189 was located on CCD module 3, which failed in quarter 4. This means that data from quarters 8, 12, and 16 are not available for this target.  Additional short-cadence (SC; around 1 minute cadence) data of \189 are available from quarter 4 up to quarter 7. For \686, one third of quarter 8 has SC data. In total, \kepler\ observed 36 transits of \189 and 26 of \686.

We recovered the uncorrected light curves issued from the photometric analysis (PA) module of the \kepler\ pipeline from the MAST archive\footnote{Mikulski Archive for Space Telescopes: \url{http://archive.stsci.edu/kepler/data_search/search.php}}. The LC light curves are plotted in Fig.~\ref{fig.rawlcs}. For the transit modeling  we only used the fragments of the light curve around each transit, The data were normalized with a quadratic function whose coefficients were fit to the out-of-transit part. This ensures a reasonably flat out-of-transit curve without involving complicated filtering methods that risk modifying the transit shape. For \189\ only LC data were used because the sampling of the transit, and in particular of the ingress and egress phases, was deemed sufficient. On the other hand, \686\  exhibits at most two transits per quarter, and only one in most of them. Therefore the LC data of \686\ are not enough to sample the ingress and egress phases of the transit adequately. Because this could lead to incorrectly determined transit parameters (Bonomo et al., submitted) the SC transit observed in quarter 8 was used instead of its LC version. A sigma-clipping was performed to reject outliers, taking into account the total number of points in each light curve. The contamination by nearby stars was corrected for using the crowding values from the \kepler\ archive. The final light curve of \189\ has 1310 points; for \686\ there are 436 points in the LC light curves and 585 points in the SC ones.

In Sect.~\ref{sect.starparam} we employ the corrected pre-search data conditioning (PDC) \kepler\ light curve to search for out-of-transit modulations due to stellar spots. These data are also available from the MAST archive.

\subsection{SOPHIE radial velocities}

SOPHIE is a high-resolution \emph{echelle} spectrograph fiber-fed from the Cassegrain focus of the 1.93-m telescope at the Haute-Provence Observatory (OHP) in France. It is installed in a temperature-stabilized environment, and the dispersive elements are kept at constant pressure to provide high-precision radial velocities \citep{perruchot2008}. \sophie\ spectra cover most of the visible wavelength range in 39 spectral orders. Since June 2011, a new fiber scrambler has provided a significant improvement of the spectrograph illumination stability, leading to a precision gain of a factor $\sim6$  \citep{perruchot2011, bouchy2013}. All the observations presented here were performed with the new fiber scrambler.

Observations were performed in high-efficiency mode, with a resolving power $\lambda/\Delta\lambda\sim 40\,000$. One of the two available optical fibers records the stellar spectrum, while the other, located 2 arcmin away, records the background sky spectrum, which can be used to correct for scattered light entering the star fiber \citep[e.g.,][]{barge2008,pollacco2008,santerne2011}. Only one spectrum of \686 needed to be corrected for contamination by the moonlight.

\189 was observed ten times between July 3 2011 and November 10 2011. Exposure times were varied between one hour for the first measurements to around 12 minutes, when it became clear that the precision required to characterize the companion orbit was achieved in shorter exposures. Therefore, the signal-to-noise ratio (S/N) is also highly variable between spectra: from 9.0 up to 18.0. In these cases, a systematic effect in the radial velocities due to the charge transfer inefficiency (CTI) in SOPHIE's CCD must be corrected for. This was made using the empirical correction described in \citet{santerne2012}. \686 was observed seven times between July 9 2011 and November 7 2011. Similarly to \189, the exposure times range between 10 minutes and one hour, and a similar correction for the CTI effect was needed.

The spectra were reduced and extracted using the \sophie\ pipeline \citep{bouchy2009}, and the resulting wavelength-calibrated 2D spectra were correlated using a numerical binary mask corresponding to spectral type G2 (for \686) and K5 (for \189) to obtain the radial velocity measurement \citep{baranne1996, pepe2002}. For faint targets such as these, the orders at the edge of the wavelength range are usually too noisy, and adding them in the average cross-correlation function degrades the precision of the measurements. Therefore, the average cross-correlation function (CCF) was computed discarding the 11 bluest orders in the spectrum. The radial velocity (RV) time series show variations with amplitudes of a few \kms, which indicates that the transiting candidates are not planetary objects. If blended binaries can be excluded (see below), the companions are instead in the regime of massive brown dwarfs or very low-mass stars.

\begin{figure}
\begin{center}
\includegraphics[width=0.7\columnwidth]{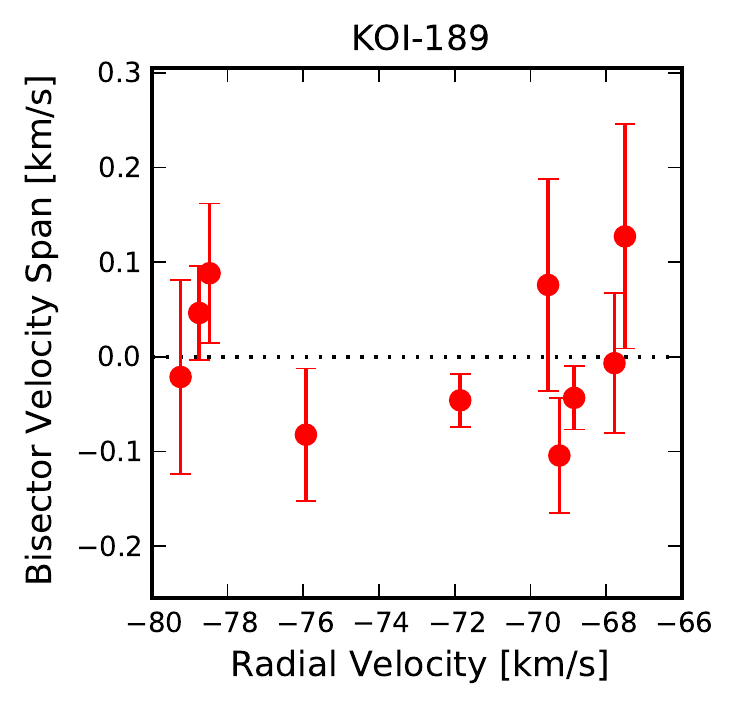}
\includegraphics[width=0.7\columnwidth]{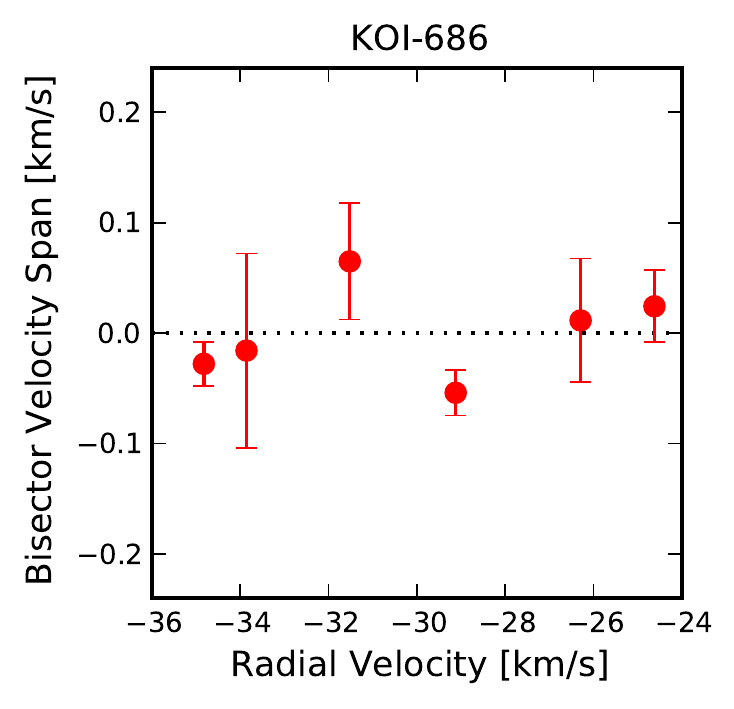}
\caption{Bisector velocity span as a function of radial velocity. The measurement of \686\ affected by moonlight contamination is not plotted. The error in the radial velocity is smaller than the size of the points. The y-axis scale is 25 times smaller than the scale of the x-axis. \label{fig.bisectors}}
\end{center}
\end{figure}

The SOPHIE pipeline also measures the FWHM of the CCF and its bisector velocity span \citep{queloz2001}. The former can be used to estimate the projected rotational velocity of the star using the calibrations presented in \citet{boisse2010} (see Sect.\ \ref{sect.starparam}); the latter measures the deformation of the CCF, which might reveal additional sets of spectral lines blended with the main spectrum or magnetic activity features in the surface of the star \citep[see][and Santerne et al. 2014, in prep., for a thorough discussion on RV diagnostics]{figueira2013}. The bisector velocity span is plotted as a function of the radial velocity in Fig.~\ref{fig.bisectors}, where the scale of the bisector axis is 25 times smaller than that of the radial velocity. It is clear that the bisector does not exhibit any variability at the scale of the radial velocity variations, and that no correlation exists between both observables. This is confirmed by the Pearson correlation coefficients:  $r = -0.31 \pm 0.34$ for \189, and $r = 0.20\pm0.49$ for \686, which are compatible with zero, and where the reported errors are obtained by a bootstrap procedure. Although the lack of correlation between the RV and bisector variations is not sufficient to rule out all possible false positives by itself, it has been argued that if additionally transits are U-shaped and clear Keplerian variations in phase with the photometric variations are detected, roughly all false-positive scenarios involving a blended binary system can be rejected \citep{bakos2012}. We therefore conclude that the radial velocity variations are produced by companions to the host star with a negligible luminosity ratio, and not by blended eclipsing binaries.

The complete radial velocity datasets are reported in Tables~\ref{table.rv189} and \ref{table.rv686} and plotted in Figs.~\ref{fig.results189} and \ref{fig.results686}. The uncertainties include the photon noise error, estimated with the method by \citet{bouchy2001} and the error in the wavelength calibration ($\sim 2$ \ms).

\onltab{
\begin{table}[t]
\caption{Radial velocity measurements for \189, obtained with the K5 mask. \label{table.rv189}}
\begin{tabular}{l l l c c c}
\hline
\hline
\noalign{\smallskip}
BJD     & RV          & $\sigma_{RV}$ & BVS\tablefootmark{a}  & Exp.\ time & S/N/pix    \\
-2 450 000  &(\kms) & (\kms) & (\ms) & (s)       & (550 nm)\\
\hline
\noalign{\smallskip}
55746.4495 &  -78.748 & 0.024 & 0.046  & 2926 &11\\
55763.5061 &  -68.845 & 0.016 & -0.043 & 3600 &17\\
55768.4028 &  -71.854 & 0.013 &  -0.046  & 3600 &18\\
55773.4533 &  -75.930 & 0.034 &  -0.082 & 2183 &12\\
55776.5059 &  -78.475 & 0.036 &  0.088 & 1200 &9\\
55785.4191 &  -69.234 & 0.030 &  -0.104 & 1800 &14\\
55788.5163 &  -67.503 & 0.059 &  0.127 & 810  &10\\
55791.4151 &  -67.779 & 0.037 &  -0.007 & 900  &11\\
55809.3730 &  -79.241 & 0.051 &  -0.021 & 1160 &10\\
55876.2423 &  -69.533 & 0.056 &  0.076 & 744  &10\\
\noalign{\smallskip}
\hline 
\end{tabular}
\tablefoot{
\tablefoottext{a}{Bisector velocity span. The associated error is twice the RV error.}
}
\end{table}
}

\onltab{
\begin{table}[t]
\caption{Radial velocity measurements for \686 \label{table.rv686}}
\begin{tabular}{l l l c c c}
\hline
\hline
\noalign{\smallskip}
BJD     & RV          & $\sigma_{RV}$ & BVS\tablefootmark{a}  & Exp.\ time & S/N/pix    \\
-2 450 000  &(\kms) & (\kms) & (\kms) & (s)       & (550 nm)\\
\hline
\noalign{\smallskip}
55752.549410 & -34.822 & 0.010 &  -0.028 & 3600 & 27\\
55776.472340 & -29.120 & 0.011 &  -0.054 & 3600 & 27\\
55785.446050 & -24.621 & 0.017 &  0.024 & 1800 & 18\\
55791.428390 & -26.292 & 0.029 &  0.012 & 600   & 11\\
55801.574040 & -35.442 & 0.066 &  0.109 & 600   & 10\\
55810.432690 & -33.857 & 0.045 &  -0.016 & 1038 & 7\\
55873.277190 & -31.519 & 0.027 &  0.065 & 900   & 12\\
\noalign{\smallskip}
\hline 
\end{tabular}
\tablefoot{
\tablefoottext{a}{Bisector velocity span. The associated error is twice the RV error.}
}
\end{table}
}

\subsection{Photometry}

\begin{table*}
\caption{Target characteristics and absolute photometric measurements. \label{table.obslog}}
\begin{center}
\begin{tabular}{p{1.9cm}cc}
\hline
\hline
\noalign{\smallskip}
			& \189 						& \686\\
\noalign{\smallskip}
\multicolumn{3}{l}{Target Information} \\
\hline
\noalign{\smallskip}
Kepler ID 		& 11391018 					&  7906882 		\\
2MASS ID 	& 18593119+4916011			&19472178+4338496\\
\noalign{\smallskip}
R.A. (J2000)  	&18 59 31.19 					& 19 47 21.78\\
Dec. (J2000) 	&+49\degr 16\arcmin 01\farcs2 	&+43\degr 38\arcmin 49\farcs.6\\
\noalign{\smallskip}
\kepler\ magnitude \tablefootmark{a} 		& 14.388		& 14.545\\
\noalign{\smallskip}
SDSS $g^\prime$\tablefootmark{a}		& 15.17		& 14.12\\
SDSS $r^\prime$\tablefootmark{a}		& 14.33		& 13.52\\
SDSS $i^\prime$\tablefootmark{a}		& 14.05		& 13.35\\
SDSS $z^\prime$\tablefootmark{a}		& 13.92		& 13.26\\
\noalign{\smallskip}
2MASS-J \tablefootmark{b} 			& $12.895\pm0.025$ &$12.270 \pm 0.020$\\
2MASS-H \tablefootmark{b} 			& $12.377\pm0.021$ &$ 11.932\pm 0.018$\\
2MASS-Ks \tablefootmark{b} 			& $12.288\pm0.025$ &$ 11.847\pm 0.019$\\
\noalign{\smallskip}
WISE-W1 \tablefootmark{c} 			& $12.229\pm0.024$ &$ 11.789 \pm 0.023$\\
WISE-W2 \tablefootmark{c} 			& $12.317\pm0.022$ & $11.844 \pm 0.021$\\
\hline
\hline
\end{tabular}
\end{center}
\tablefoot{
\tablefoottext{a}{From Kepler Input Catalogue.}
\tablefoottext{b}{\citet{2MASS}}
\tablefoottext{c}{\citet{cutri2012}}
}
\end{table*}

In addition, we recovered infrared photometric measurements of these two targets from the 2MASS \citep[][]{2MASS} and from the WISE space mission \citep{wright2010}. In the optical part of the spectrum, measurements in the SDSS photometric bands are available in the KIC. The uncertainties in these bands are stated to be between 0.03 and 0.04 mag. We conservatively decided to use 0.04 mag for all four available SDSS bands. We list the magnitudes for the two targets in Table~\ref{table.obslog}. These data were employed to constrain the parameters of the host star and its distance by fitting the spectral energy distribution (see Sect.\ \ref{sect.modeling}). 

\section{Primary star  parameters \label{sect.starparam}}

\begin{table*}
\caption{Stellar parameters of \189 for different stellar evolution models.\label{table.tracks189}}
\begin{tabular}{l|c|ccc}
\hline
\hline
Parameter								& Spectroscopy 	& Dartmouth 		& Parsec 			& StarEvol \\
\hline
Effective temperature, $T_{\mathrm{eff}}$[K]	&$4850 \pm 100 $ 	&$4949 \pm 42$ 	&$4955\pm42$ 	&$4952\pm36$ \\
Metallicity, $[\rm{Fe/H}]$ [dex]				&$-0.07 \pm 0.12$	&$-0.13 \pm 0.13$	&$-0.11\pm0.14$	&$-0.114\pm0.056$\\
Surface gravity, log\,$g$ [cgs]				&$4.6\pm0.12$		&$4.591 \pm 0.014$ 	&$4.582\pm0.016$	&$4.5920\pm0.0069$\\
Stellar density, $\rho_{\star}$ [$\rho_\odot$] 	& --				&$1.924 \pm 0.054$	&$1.926\pm0.056$	&$1.936\pm0.052$\\
Star mass, $M_\star$ [\Msun]				& --				&$0.774 \pm 0.045$ &$0.730\pm0.059$	&$0.775^{+0.025}_{-0.036}$\\
Star radius, $R_\star$ [\Rsun]				& --				&$0.738 \pm 0.015$ &$0.724\pm0.019$	&$0.735\pm0.016$\\
$\log_{10}\tau$ [yr]						& --				&$9.79^{+0.22}_{-0.43}$&$10.05^{+0.19}_{-0.32}$	&$9.79\pm0.18$\\
Distance [pc]							& --				&$424 \pm 12$		&$417\pm14$		&$423\pm12$\\
Companion mass, $M_c$ [\Msun]			& --                		&${0.0751\pm0.0028}$&${0.0724\pm0.0038}$&${0.0752\pm0.0023}$\\
Companion radius, $R_c$[\Rsun]			& --				&${0.1032\pm0.0021}$&${0.1013\pm0.0025}$&${0.1028\pm0.0025}$\\
\hline
\end{tabular}
\end{table*}

\begin{table*}
\caption{Stellar parameters of \686 for different stellar evolution models. \label{table.tracks686}}
\begin{tabular}{l|c|ccc}
\hline
\hline
Parameter								& Spectroscopy & Dartmouth & Parsec & StarEvol \\
\hline
Effective temperature, $T_{\mathrm{eff}}$[K]	&${5750\pm120}$ 	&${5834\pm100}$		&${5834\pm100}$		&${5836\pm94}$\\
Metallicity, $[\rm{Fe/H}]$ [dex]				&${0.02\pm0.12}$	&${-0.06\pm0.13}$		&${-0.06\pm0.13}$		&${-0.06\pm0.12}$\\
Surface gravity, log\,$g$ [cgs]				&${4.5\pm0.15}$	&${4.399\pm0.015}$	&${4.391\pm0.016}$	&${4.400\pm0.017}$\\
Stellar density, $\rho_{\star}$ [$\rho_\odot$] 	& --					&${0.874\pm0.031}$	&${0.874\pm0.31}$	&${0.873\pm0.032}$\\
Star mass, $M_\star$ [\Msun]				& --					&${0.995\pm0.057}$	&${0.950\pm0.070}$	&${1.002\pm0.075}$\\
Star radius, $R_\star$ [\Rsun]				& --					&${1.044\pm0.025}$	&${1.029\pm0.029}$	&${1.047\pm0.029}$\\
$\log_{10}\tau$ [yr]						& --					&${9.75^{+0.14}_{-0.20}}$	&${9.86\pm0.20}$		&${9.77^{+0.18}_{-0.28}}$\\
Distance [pc]							& --					&${531\pm17}$			&${523\pm19}$		&${532\pm19}$\\
Companion mass, $M_c$ [\Msun]			& --                			&${0.0994\pm0.0038}$		&${0.0965\pm0.0047}$	&${0.1000\pm0.0049}$\\
Companion radius, $R_c$[\Rsun]			& --					&${0.1254\pm0.0031}$		&${0.1236\pm0.0036}$	&${0.1258\pm0.0036}$\\
\hline
\end{tabular}
\end{table*}

The co-added SOPHIE spectra were used to measure the atmospheric parameters of the transit host using the method described by \citet{deleuil2012}, which yields the values listed in the second column of Tables~\ref{table.tracks189} and~\ref{table.tracks686}.

The \kepler\ PDC out-of-transit light curves were used in an attempt to estimate the rotational period of the primary stars, and their age by gyrochronology. For \189, a generalized Lomb-Scargle periodogram \citep[GLS, ][]{ZechmeisterKurster2009} reveals a significant period at P = $30.45 \pm 0.35$ days, with a peak-to-peak amplitude of $\sim 0.2$\%. In agreement, \citet{walkowiczbasri2013} reported a rotation period of P = $33.26 \pm 3.44$ days for this star, based on the analysis of quarter 9 data alone. On the other hand, \citet{mcquillan2013b} did not find a significant period for this star from the auto-correlation function on quarter 3-14 data. This periodicity could be interpreted as the rotational period because it is present identically in the first and second halves of the light curve. However, it is known that the \kepler\ pipeline attenuates astrophysical signals with periods longer than about ten days, at least for amplitudes around 0.1 \%\footnote{See PDC Data Release 21, Sect.\ 3.1.3, \url{https://archive.stsci.edu/kepler/release_notes/release_notes21/DataRelease_21_20130508.pdf}}. Moreover, this period is also similar to the period of the transiting candidate. Because the star is not expected to be synchronized (see Sect. \ref{sect.radius}), this raises further suspicion about the nature of this periodicity. We decided to remain conservative and do not claim the detection of the rotational period of the primary star of the \189\ system from the \kepler\ light curve.

On the other hand, the projected rotational velocity measured spectroscopically is $v \sin i = 2.5 \pm 1.5$ \kms. The FWHM of the SOPHIE CCF can also be used to estimate roughly the projected rotational velocity of the star using the calibrations by \citet{boisse2010}, which gives $4.4 \pm 1.0$ \kms, roughly in agreement within the errors.

The PDC light curve of \686\ is clearly variable, with the highest peak in a GLS periodogram at P = $13.66\pm0.07$ days. There is power at a number of other peaks between 11 and 21 days, however. When the light curve is split into three 500-day sections, the highest peak in the GLS analysis changes:  P =$19.9\pm0.4$ days for $BJD < 2455400$,  P = $15.5\pm0.24$ days for $2455400 < BJD < 2455900$, and P = $13.66\pm0.17$ days for  $BJD > 2455900$. Although this could be reminiscent of differential rotation and migrating spots, the amplitude of the modulation remains roughly constant, and a similar variability is detected in some of the cotrending basis vectors that are not corrected for by the \kepler\ pipeline \citep[see][]{stumpe2012, smith2012}. In summary, we do not claim the detection of a periodic light-curve modulation that could be interpreted as caused by stellar spots. We are therefore unable to obtain an estimate of the stellar age and the rotational period. We note that neither \citet{walkowiczbasri2013} nor \citet{mcquillan2013b} reported a significant rotational period either. The projected rotational velocity obtained from the FWHM of the CCF is $v \sin i = 3.5 \pm 1.0$ \kms, in agreement with the value obtained from the spectroscopic analysis, $v \sin i = 3.0 \pm 1.5$ \kms.

\section{Data analysis \label{sect.modeling}}

\subsection{Data modeling}
The \kepler\ light curves and SOPHIE radial velocities were fitted together with the broadband absolute photometry data. The model consists of a star orbited by a dark companion of a given mass and radius. We employed the modeling procedure and Markov chain Monte Carlo (MCMC) module from the PASTIS package \citep{diaz2014}. We refer to this article for details on the modeling of the star, the companion, and the MCMC algorithm. 

The host star was parametrized using its mean density, metallicity, and effective temperature. The atmospheric analysis described in Sect.\ \ref{sect.starparam} provides priors for these three parameters. Stellar evolution models were used to obtain self-consistently the stellar luminosity and radius, which were used to scale the stellar atmospheric model interpolated from the grid of \citet{allard2012}. The scaled spectrum was corrected for the extinction and used to compute the model of the observed absolute magnitudes. The limb-darkening coefficients were obtained self-consistently from the table of \citet{claret2011} by interpolation of the stellar parameters.

To estimate the contribution of systematic errors, three different stellar evolution models were used. These are the Dartmouth \citep{dotter2008}, Parsec \citep{bressan2012}, and StarEvol (Palacios, priv. comm.) models. Because the transit depth and radial-velocity semi-amplitude can be measured with a very high precision for these objects, the error budget of their masses and radii will be dominated by the systematic errors of the primary mass and radius. 

The priors used for the MCMC algorithm are listed in Table~\ref{table.priors}. Basically, we used non-informative priors --that is, uniform or Jeffreys distributions \citep[see][]{gregory}-- for all parameters except for the stellar parameters mentioned above, and the ephemerides parameters, P and Tc, for which we used the values provided by \citet{batalha2013}, but the width of the distribution was increased by an order of magnitude to avoid biasing our results. To explore parameter space as thoroughly as possible, we ran 30 chains for each set of stellar tracks. The chains were started at random points drawn from the prior distribution and were run for $1\times10^{6}$ iterations. In all cases, most of the chains converged to indistinguishable distributions, according to the Kolmogorov Smirnov test applied on the posterior distribution. This supports the assumption that the posterior distribution is unimodal and that the global maximum was found. The chains were thinned using their correlation length, and merged to form the final sample of the parameter posterior distribution. In all cases we obtained more than 50,000 independent samples from the posterior. The model parameters and their uncertainties were inferred from a combination of equal-sized samples of the posterior for each set of stellar evolutionary tracks. In this way, the systematic errors introduced by the models were taken into account.
\onltab{
\begin{table*}[t]
  \centering 
  \caption{Priors of fitted parameters. \label{table.priors}}
  \label{tableMCMC}
\begin{tabular}{llcc}
\hline
\hline
\noalign{\smallskip}
					&			& \centering \189														& \686 \\
\noalign{\smallskip}

{\bf Parameters	}		&{\bf Units}	&\multicolumn{2}{c}{\bf Prior}												\\
\hline
\noalign{\smallskip}
\multicolumn{3}{l}{\it Orbital Parameters}\\
\hline
\noalign{\smallskip}
Period, $P$						& days				& $\mathrm{N}(30.360446; 1.4\times10^{-4})$		& $\mathrm{N}(52.51357; 1.6\times10^{-4})$	\\
Midtransit time, $T_c$				& BJD\_UTC - 2,454,000	&  $\mathrm{N}(981.09095; 1.5\times10^{-3})$		& $\mathrm{N}(1004.6737; 1.1\times10^{-3})$	\\
Eccentricity, $e$					& --					& \multicolumn{2}{c}{$\mathrm{U}(0.0; 1.0)$}\\
Argument of periastron, $\omega$		& deg 				& \multicolumn{2}{c}{$\mathrm{U}(0.0; 360.0)$}\\	
Impact parameter, $b$				& --					& \multicolumn{2}{c}{$\mathrm{U}(0.0; 1.0)$}\\	
\noalign{\smallskip}

\multicolumn{3}{l}{\it Stellar Parameters}\\
\hline
\noalign{\smallskip}
Effective temperature, $T_\mathrm{eff}$	& K					& $\mathrm{N}(4850; 100)$					& ${\mathrm{N}(5750; 120)}$ \\
Metallicity, [Fe/H]					& dex				& $\mathrm{N}(-0.07; 0.18)$					& ${\mathrm{N}(0.02; 0.12)}$\\
Stellar density, $\rho_*$ 				& [$\rho_\odot$] 		& $\mathrm{N}(1.79; 0.30)$					& ${\mathrm{N}(0.91; 0.28)}$\\
Distance, $d$						& pc					&  \multicolumn{2}{c}{$\mathrm{U}(100; 2000)$}\\

\noalign{\smallskip}

\multicolumn{3}{l}{\it System Parameters}\\
\hline
\noalign{\smallskip}
Radial-velocity semi-amplitude, $K$		& [km s$^{-1}$]			& \multicolumn{2}{c}{$\mathrm{U}(3.0; 9.0)$}\\
Radius ratio, $k  = Rc/Rs$ 			& -- 					&  \multicolumn{2}{c}{$\mathrm{J}(0.01; 0.5)$}\\	 
Center-of-mass velocity, $\gamma$ 		& [km s$^{-2}$] 		& $\mathrm{U}(-78.0, -70.0)$					& $\mathrm{U}(-32.8;  -28.8)$\\
\noalign{\smallskip}

\multicolumn{3}{l}{\it Data Parameters}\\
\hline
\noalign{\smallskip}
Out-of-transit flux, $f_\mathrm{oot}$ 		& -- 					&  \multicolumn{2}{c}{$\mathrm{U}(0.999;  1.001)$}\\	
Kepler long cadence jitter, $\sigma_\mathrm{LC}$ 	&[ppm]		& \multicolumn{2}{c}{$\mathrm{U}(0.0;  800.0)$}\\
Kepler long cadence jitter, $\sigma_\mathrm{SC}$ 	&[ppm]		& $\mathrm{U}(0.0;  4000.0)$ 	& 	--\\
SOPHIE jitter, $\sigma_\mathrm{SOPHIE}$		&[\ms]		& \multicolumn{2}{c}{$\mathrm{U}(0.0;  120.0)$}\\
SED jitter, $\sigma_\mathrm{SED}$				&[mag]		& \multicolumn{2}{c}{$\mathrm{U}(0.0;  1.0)$}\\
\noalign{\smallskip}

\hline
\hline
\end{tabular}
\tablefoot{\\
$\mathrm{U}(x_{min};  x_{max})$: uniform distribution between $x_{min}$ and $x_{max}$.\\
$\mathrm{J}(x_{min};  x_{max})$: Jeffreys (log-flat) distribution between $x_{min}$ and $x_{max}$.\\
$\mathrm{N}(\mu; \sigma)$: normal distribution with mean $\mu$ and standard deviation $\sigma$.\\
$\mathrm{AN}(\mu; \sigma_-; \sigma_+)$: asymmetric normal distribution, with different widths at each side of mean value.
}

\end{table*}
}

\section{Results \label{sect.results}}

In Tables~\ref{table.tracks189} and~\ref{table.tracks686} we list the median and 68.3\% confidence intervals for selected parameters for each set of stellar tracks used. The stellar atmospheric parameters deduced using each of the three tracks agree with each other. They also agree well with the stellar atmospheric parameters obtained from the spectroscopic analysis, although the $T_\mathrm{eff}$ from the combined fit is systematically found to be pushed to the high end of the prior distribution. We repeated the analysis for the Dartmouth tracks using different, less constraining priors for the stellar parameters: a 1000-K-wide normal distribution for the $T_\mathrm{eff}$, a uniform distribution between 0.8 and 3.0 times the solar density and a normal distribution with a width of 1.8 dex for the metallicity. The resulting parameters agree within 1 $\sigma$ with those reported in Tables~\ref{table.tracks189} and \ref{table.tracks686}, and the posterior distributions have a similar shape and width. This shows that the priors are not restricting the Markov chains from attaining different values, but is indicative instead of a systematic difference between the  $T_\mathrm{eff}$ obtained from the spectroscopic analysis and the $T_\mathrm{eff}$ obtained by fitting broadband photometry, as previously reported in the literature \citep[e.g.,][]{pinsonneault2012, huber2014}.

As expected, the surface gravity of the primary star derived from the MCMC by the bulk stellar density obtained from the transit shape is much more precise than that obtained from the spectrum. The PARSEC stellar tracks provide marginally smaller and less massive primaries, which results in a smaller, less massive companions, but the difference only amounts to 0.5-$\sigma$ both in $R_c$ and $M_c$. Therefore, we conclude that the result agrees for all three models. The ages of the systems are unconstrained because of the relatively small stellar mass.

The results of the combined chains --obtained as described at the end of the previous section-- are presented in Table~\ref{table.Params}. The final error in the masses and radii of the companions is between 5\% and 8\% larger than the errors inferred from individual tracks.

\begin{figure}[t]
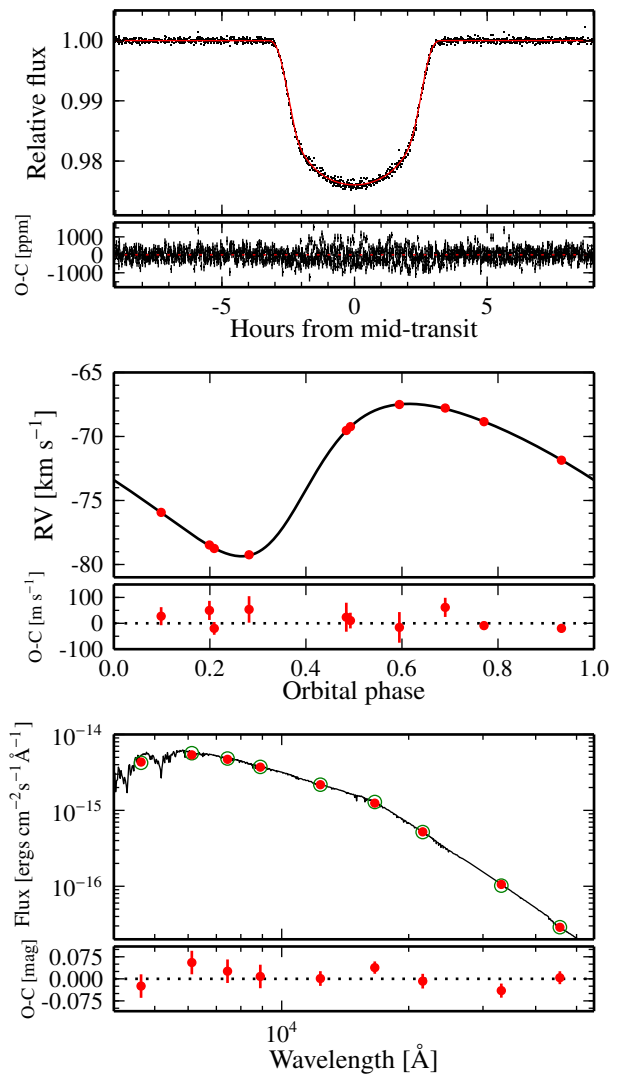

\input{KOI-189_keplerLC_phase.pgf}
\input{KOI-189_SOPHIE_phase.pgf}
\input{KOI-189_SED.pgf}
\caption{Data and best-fit model for \189. From top to bottom: phase-folded  \kepler\ LC light curve, SOPHIE RV curve, and spectral energy distribution. The \kepler\ data are corrected for contamination and normalized to the out-of-transit flux listed in Table~\ref{table.Params}. The model of the \kepler\ LC data is shown binned down to the data sampling rate. For the SED, the best-fit spectrum is plotted as a solid black curve, and the integrated fluxes in the photometric bands are plotted as open circles. \label{fig.results189}}
\end{figure}

\begin{figure*}[t]
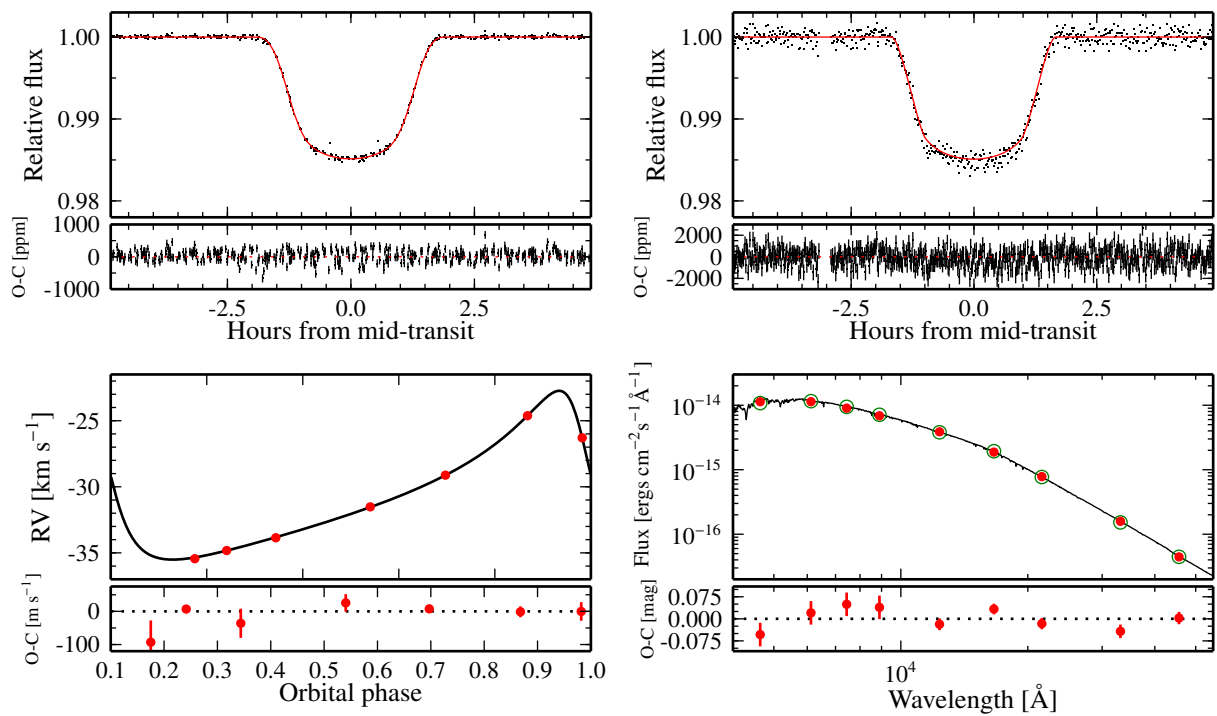

\begin{center}
\input{KOI-686_keplerLC_phase.pgf}
\input{KOI-686_keplerSC_phase.pgf}
\input{KOI-686_SOPHIE_phase.pgf}
\input{KOI-686_SED.pgf}
\caption{Data and best-fit model for \686. Top row: phase-folded  \kepler\ LC (left) and SC (right) light curves. Bottom row: SOPHIE RV curve (left), and spectral energy distribution (right). The \kepler\ data are corrected for contamination and normalized to the out-of-transit flux listed in Table~\ref{table.Params}. The model of the \kepler\ LC data is shown binned down to the data sampling rate. For the SED, the best-fit spectrum is plotted as a solid black curve, and the integrated fluxes in the photometric bands are plotted as open circles. \label{fig.results686}}
\end{center}
\end{figure*}

\subsection{\189}
The MCMC analysis of the available data indicates that \189\ is a K-type dwarf star with roughly solar metallicity and an inferred mass $M_\star = 0.763 \pm 0.054$ \Msun, and radius $R_\star = 0.731 \pm 0.018$ \Rsun. 

The transiting companion has a radius of $R_c = 0.1025 \pm 0.0024$ \Rsun, and a mass $M_c = 0.0745 \pm 0.0033$ \Msun. Its density is, therefore, $79.4 \pm 3.7$ times the density of Jupiter, and its surface gravity $\log g_c  = 5.291\pm0.013$. In Fig.~\ref{fig.results189} we present the maximum-posterior model of the system together with the data points and the residuals. The position of the transiting companion in a mass radius diagram (Fig.~\ref{massradius}) agrees well with the model predictions for ages slightly below 1 Gyr. Although the age estimated based on the stellar tracks is closer to 6 Gyr, 1 Gyr is within the 85\% confidence interval. In any case, the radius of the companion is larger than predicted by models for ages above 1 Gyr.  With respect to the \citet{baraffe2003} models, the posterior radius anomaly, defined as the fractional difference between the measured radius and the model prediction, is $5.1 \pm 1.3$ \% for the 1 Gyr isochrone and $14.3 \pm 3.8$ \% for the 5 Gyr isochrone. An informative way to report this discrepancy is by employing the highest-density interval (HDI) of the posterior distribution of the radius anomaly. The q\%-HDI of a distribution is defined as the (possibly discontinuous) interval containing q\% of the distribution mass, and such that all points within the interval have a higher density than those outside. By definition, the HDI contains the mode of the distribution. In this case, the 95\%-HDI of the radius anomaly of \189 B is [2.6\%; 7.7\%] for the 1 Gyr isochrone and [7.7\%; 20.0\%] for the 5 Gyr isochrone, which emphasizes the disagreement between observations and model.

\subsubsection*{Secondary-eclipse search}
The impact parameter at superior conjunction indicates that the system undergoes secondary eclipses. We searched for periodic dips in the Kepler light curve at the expected orbital phase using a box model with three parameters: central phase, duration, and depth. The posteriors estimated from the MCMC algorithm were used as priors for the phase and duration parameters. For the depth, we used a Jeffreys prior between $10^{-12}$ and $10^{-4}$. We checked that the limits used in the Jeffreys distribution did not change the computation significantly. We adopted a Gaussian model for the errors, with variance $\sigma^2  = \sigma_i^2 + \sigma_J^2$, where $\sigma_i$ are the internal errors provided by the Kepler pipeline and $\sigma_J$ is the mean LC jitter reported in Table~\ref{table.Params} and obtained for the transit modeling. The Bayes factor between the box model and the null hypothesis (no secondary eclipse) is close to 1. Furthermore, we searched in the same manner at 100 other orbital phases linearly sampled between one fourth of a period before and after the expected time. The Bayes factor ranges between 0.96 and 1.17, showing that the light curve around the orbital phase where the secondary eclipse is expected is actually typical.

It is interesting to note that if the data model is chosen as a Gaussian distribution with variance $\sigma_i^2$, meaning, without the additional jitter term, then the Bayes factor in favor of a secondary eclipse is close to 10, which could have been interpreted as a positive detection \citep{kassraftery1995}. However, searching at other phases yields many occurrences of Bayes factors at similar and higher values. This shows that systematics in the Kepler light curve and stellar noise can easily mimic secondary eclipses of the correct depth if the data modeling is not correct. The jitter term is a basic method to deal with this problem, but to fully exploit these data, a more sophisticated noise model is needed \citep[see, e.g.,][]{parviainen2014}.

The non detection of the secondary eclipse was expected from theoretical isochrones. According to \citet{baraffe2003}, the absolute magnitude of the transiting companion in the $r$ band is around 17.8 (assuming an age of 5 Gyr), which leads to a magnitude difference with the host star $\Delta_r = 11.6$, and an expected secondary depth of 23.5 ppm, well below the dispersion of the available data and of the measured jitter value. On the other hand, in the J band the absolute magnitude is 12.2, and then $\Delta_J = 7.4$, and the secondary depth is 1090 ppm. 

\subsection{\686}
\686 is a G-type star, with a mass $M_\star = 0.983 \pm 0.074$ \Msun, and radius $R_\star = 1.040 \pm 0.030$ \Rsun. The companion has a mass $M = 0.0987 \pm 0.0049$ \Msun\ and a radius $R = 0.1250 \pm 0.0038$ \Rsun. The maximum-posterior model and the residuals are plotted in Fig.~\ref{fig.results686} .

Unlike \189, the \686\ system is not expected to exhibit secondary eclipses, since the impact parameter at superior conjunction is 2.05. The inverse situation, in which the primary eclipse of low-mass stellar companions is missing is known to cause false-positive detection in transiting planet surveys \citep{santerne2013}.

From an analysis of Q0-Q6 data of \686\ combined with a high-resolution Keck spectrum of the star, \citet{dawsonjohnson2012} estimated the eccentricity of the orbit of the transiting companion thanks to the difference between the transit-determined stellar density and that obtained from the spectroscopic analysis of the Keck spectrum (the so-called photoeccentric effect). The authors measured an eccentricity $e = 0.62^{+0.18}_{-0.14}$. This value agrees with the one presented here, although the RV determination is much more precise, as expected for cases where the RV signal is clearly detected, yielding an error bar about 40 times smaller. This shows that the photoeccentric effect is a valuable tool for identifying eccentric transiting candidates, but that RV remains an important asset for the precise characterization of the system, especially for long-period orbits, for which the transits are usually scarce.

\begin{figure}
\begin{center}
\input{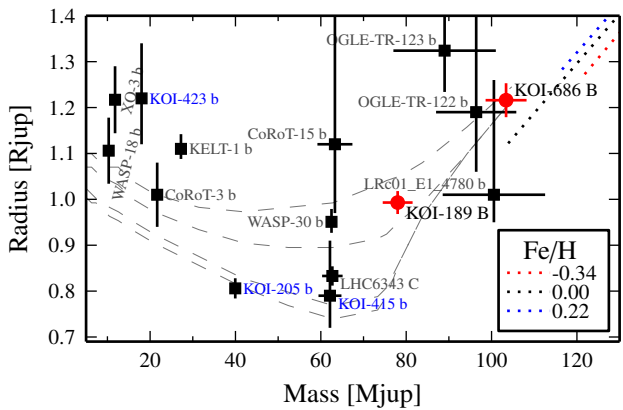}
\caption{Mass-radius diagram for massive planets, brown dwarfs, and very low-mass stars transiting primaries with $M_s > 0.5$ \Msun. The dashed lines are the isochrones of \citet{baraffe2003} for ages of (from top to bottom) 0.5, 1, 5, and 10 Gyr. The dotted lines are the 5 Gyr Dartmouth isochrones \citep[2012 version,][]{feidenchaboyer2014} for the extreme allowed values for \686, and for solar metallicity. The objects presented in this paper are plotted in red; other objects from the SOPHIE follow-up of \kepler\ candidates are labeled in blue\label{massradius}.}
\end{center}
\end{figure}

The radius of \686\ b is obtained here with a precision close to 3\%. Neglecting possible remaining systematic errors introduced by the use of the stellar tracks to obtain the mass of the primary star, this object can be used for a comparison with stellar evolution models. The models by \citet{baraffe2003} predict a radius in agreement with observations: a consistent radius value is within the 80\%-HDI of the posterior radius anomaly for both the 1 Gyr and 5 Gyr isochrones. In this case, a comparison with the newest Dartmouth isochrones \citep{feidenchaboyer2013, feidenchaboyer2014}, which include various effects due to magnetic fields, is also possible. Using the web-based interpolation tool provided by Dartmouth\footnote{\url{http://stellar.dartmouth.edu/models/isolf_new.html}}, we computed the isochrones for the metallicity values at either end of the 95\%-HDI (-0.34, 0.22), which are plotted in Fig.~\ref{massradius}. Comparison with the solar-metallicity isochrone\footnote{To compute the posterior radius anomaly of \686 B with respect to the isochrones of Baraffe and the Dartmouth group, we have only kept the samples of the posterior within the mass range of the isochrones and assumed that the remaining samples follow the same distribution.} indicates that the radius of \686 B is also inflated with respect to this model: the posterior radius anomaly is $6.5 \pm 1.9$ (95\%-HDI = [2.8\%; 9.5\%]).

\begin{table*}[t]
\caption{Posterior parameters of the \189\ and  \686\ systems. \label{table.Params}}            
\begin{minipage}[t]{10.0cm} 
\setlength{\tabcolsep}{5.0mm}
\renewcommand{\footnoterule}{}                          
\begin{tabular}{l c c}        
\hline\hline                 
\noalign{\smallskip}
										& \centering \189						& \686 \\
\noalign{\smallskip}

\multicolumn{3}{l}{\hspace{-0.5cm} Spectroscopic parameters} \\
\hline
Effective temperature, $T_{\mathrm{eff}}$[K]   		& $4850\pm 100$  						& ${5750 \pm 120}$   \\
Surface gravity, log\,$g$ [cgs]              			& $4.6 \pm 0.12$  						& ${4.5 \pm 0.15}$ 	\\
Metallicity, $[\rm{Fe/H}]$ [dex]             			& $-0.07\pm 0.18$						& ${0.02 \pm 0.12}$  	\\
Stellar rotational velocity, {$v \sin i_\star$} [\kms] 	& $2.5 \pm1.5$    						& $3.0 \pm 1.5$    	\\
Microturbulent velocity, $v_{micro}$ [\kms]   		& $1.2 \pm 0.1$  						& $0.8 \pm 0.1$   	\smallskip\\

\multicolumn{3}{l}{\hspace{-0.5cm} Result from combined analysis of the light curve, radial velocity, and spectral energy distribution} \\
\hline
Orbital period, $P$ [days]$^{\bullet}$ 			& $30.3604467 \pm 4.6\times 10^{-6}$ 	  	&$52.5135456\pm7.7\times 10^{-6}$\\
Midtransit time, $T_\mathrm{c}$ [BJD]$^{\bullet}$     	& $2454981.09148 \pm 1.3\times 10^{-4}$ 	&$2455004.67485\pm1.3\times10^{-4}$\\
$\mathrm{cov}(P, T_\mathrm{c})$ [days$^2$]             &$-4.62\times 10^{-10}$					&$ -6.30\times 10^{-10}$\\
Orbital eccentricity, $e$$^{\bullet}$              		& $0.2746 \pm0.0037$			 		&$0.5560\pm0.0037$\\
Argument of periastron, $\omega$ [deg]$^{\bullet}$ 	& $240.38\pm 0.92$						&$60.56\pm0.74$\\
Orbit inclination, $i$ [deg]			           		& $89.697\pm0.022$  		         		&$88.494 \pm 0.044$  \smallskip \\

Impact parameter in inferior conjunction, $b$$^{\bullet}$	& $0.339\pm0.022$					&${0.714\pm0.0075}$\\
Impact parameter in superior conjunction			& $0.208\pm0.014$						&${2.055\pm0.042}$\\
Orbital semi-major axis, $a$ [AU]               		& $0.1802 \pm 0.0044$					&${0.2823\pm0.0075}$\\
Normalized semi-major axis, $a/R_{\star}$	  	& $52.78\pm0.52$						&${58.38\pm0.76}$  \smallskip \\

Radius ratio, $k=R_p/R_\star$$^{\bullet}$                	& $0.13988\pm4.0\times10^{-4}$			&${0.12017\pm3.4\times10^{-4}}$ \\
Radial velocity semi-amplitude, $K$ [\kms]$^{\bullet}$  	& $5.948\pm0.033$					&$6.400 \pm 0.037$\\
Systemic velocity, $V_{r}$ [\kms]$^{\bullet}$          		& $-72.592 \pm 0.023$     				&$-30.893 ^{+0.012}_{-0.028}$ \smallskip\\

Jitter Kepler long cadence, $\sigma_{Kepler, LC}$ [ppm] $^{\bullet}$       	&$272.4\pm9.1$	     	&${158.2 \pm 9.5}$\\
Jitter Kepler short cadence, $\sigma_{Kepler, SC}$ [ppm] $^{\bullet}$       	& ---				     	&$<310$; $<360^{\dagger}$\\
Jitter radial velocity, $\sigma_{RV}$ [\ms]$^{\bullet}$	                 		&$<74$; $<99^{\dagger}$	&$<93$; $<113^{\dagger}$\\
Jitter SED, $\sigma_{SED}$ [mag]$^{\bullet}$                              		&$0.056 \pm 0.020$		&${0.030 \pm 0.020}$ \smallskip\\

Effective temperature, $T_{\mathrm{eff}}$[K]$^{\bullet}$   				&$4952\pm40$			&${5834\pm100}$ \\
Metallicity, $[\rm{Fe/H}]$ [dex]$^{\bullet}$             					&$-0.115\pm0.099$		&${-0.06\pm0.13}$ \\
Stellar Density, $\rho_{\star}$ [$\rho_\odot$]$^{\bullet}$ 				&$1.928 \pm0.054$		&${0.874\pm0.034}$ \\
Star mass, $M_\star$ [\Msun]                           						&$0.764\pm0.051$		&${0.983\pm0.074}$ \\
Star radius, $R_\star$ [\Rsun]                         						&$0.733\pm 0.017$		&${1.040\pm0.030}$ \\
Deduced stellar surface gravity, $\log$\,$g$ [cgs]               			&$4.590\pm0.014$		&${4.397\pm0.017}$ \\
$\log_{10}\tau$ [yr]		                     							&$9.84\pm0.29$		&${9.79^{+0.15}_{-0.25}}$ \\
$\tau$ [Gyr]												&${6.9^{+6.4}_{-3.4}}$; ${[0.6, 13.9]}^\ddag$	&${6.2\pm2.8}$; ${[1.3, 12.2]}^\ddag$\\

Distance to the system, [pc]$^{\bullet}$                 					&$421\pm13$			&${529 \pm 19}$ \\
Color excess, $E_{(B-V)}$ [mag]$^{\bullet}$               				&$<0.044$; $<0.059^{\dagger}$	&${0.122 \pm 0.032}$ \smallskip\\

Companion mass, $M_c$ [\Msun]             						&$0.0745\pm0.0033$	&${0.0987\pm0.0049}$ \\
Companion radius, $R_c$[\Rsun]              						&$0.1025 \pm0.0024$	&${0.1250 \pm 0.0038}$\\
Companion mass, $M_c$ [\MJ]                						&$78.0\pm3.4$			&${103.4\pm4.8}$ \\
Companion radius, $R_c$[\RJ]                  						&$0.998 \pm0.023$		&${1.216\pm0.037}$ \\
Companion surface gravity, $\log$\,$g_{c}$ [cgs]  					&$5.287\pm0.012$		&${5.239\pm0.014}$\\ 
Companion mean density, $\rho_{c}$ [$\rho_{J}$]  					& $78.4\pm 3.3$		&${57.5\pm3.0}$ \\
Companion mean density, $\rho_{c}$ [$g\;cm^{-3}$]  					& $97.3\pm4.1$		&${71.4\pm3.7}$ \\
Companion equilibrium temperature$^b$, $T_{eq}$ [K]				&$482.1\pm4.9$ 		&${540 \pm 12}$ \smallskip\\
\hline
\hline       
\vspace{-0.5cm}
\end{tabular}
\tablefoot{
$^{\bullet}$ MCMC jump parameter.\\ 
$^{\dagger}$ 95\%- and 99\%-confidence upper limit.\\
$^\ddag$ 95\% highest density interval.\\
$^b$  $T_{eq}=T_{\mathrm{eff}}\left(1-A\right)^{1/4}\sqrt{\frac{R_\star}{2 a}}$, for fixed albedo $A=0$.\\
\Msun = $1.98842\times10^{30}$ kg; \Rsun = $6.95508\times10^8$ m; \MJ $= 1.89852\times10^{27}$ kg; \RJ $= 7.1492\times10^7$ m
}
\end{minipage}
\label{tableparam}  
\end{table*}

\section{Summary and discussion \label{sect.conclusions}}

We have presented radial velocity measurements obtained with the SOPHIE spectrograph of two \kepler\ planetary candidates. These data permit identifying the candidates as very low-mass stars, stressing the importance of these types of observations for distinguishing transiting candidates from \emph{bona fide} planets. A combined analysis of the SOPHIE radial velocities, the available absolute photometric measurements, and all the available \kepler\ photometry allowed us to obtain precise mass and radius measurements of the transiting companions, yielding the most precise determination of the mass and radius of a star with mass below 0.1 \Msun\ (\686 B), and revealing a slightly lower density than expected from theoretical models for \189 B. The position of \686 B in the mass-radius relation agrees with theoretical expectations based on the isochrones of \citet{baraffe2003}. However, the more recent, and more complete models by the Dartmouth group \citep{feidenchaboyer2013, feidenchaboyer2014} failed to reproduce the observations, and give a radius anomaly of around 6\%, not unlike other objects in the literature, such as CM Draconis \citep[][see also \citet{feidenchaboyer2012a}]{morales2009, terrien2012}. In addition, \189 B might also be a high-mass brown dwarf, since the H-burning limit of $\sim 72$ \MJ\ is contained in the 95\%-HDI of the mass posterior distribution.

Although systematic effects in the estimation of the primary mass remain a concern for these determinations, our analysis tried to take into consideration their effect by using three different stellar evolution models to go from the measured stellar atmospheric parameters to masses and radii: these are the stellar evolution tracks from the Dartmouth and Padova groups, and the STAREVOL tracks. Although the stellar parameters obtained with the different tracks agree, the error on the companion radius increases by almost 10\% when the results from the different models are combined. A source of error in the companion radius not considered here is the uncertainty in the crowding factor reported in the MAST. This value is given without an error in the database, but with three significant digits. Assuming the error is in the second digit, then its relative error is around 1\%, which is smaller than the 3\% uncertainty in the stellar radius, which dominates the error budget of $R_c$ for these large objects for which the transit depth is known very precisely.

\subsection{Inflated radius of \189 b \label{sect.radius}}

Even if the discrepancy between the measured radius of \189 b and the 5\,Gyr isochrone of \citet{baraffe2003} is small, it is significant and confirms the radius anomaly of low-mass stars down to the very bottom of the main sequence \citep[see][and references therein]{torres2013}

Because a very large amount of energy is needed to inflate such a massive brown dwarf or very low-mass star, both the irradiation by the primary and tidal heating are found to be negligible. Indeed, at this age, a 78\,\MJ\ object is expected to release an internal flux corresponding to an effective temperature of $1900-2000$\,K. To affect the evolution, any heating mechanism should yield a comparable flux. Because of the wide orbit, the irradiation of the primary and the tidal dissipation due to the eccentricity yield an equilibrium temperature of ~480\,K and 10-20\,K\footnote{The tidal energy flux is computed with the tidal model of Leconte et al. (2010) assuming a Jupiter-like dissipation efficiency.}, which is orders of magnitude too low (the irradiation represents 0.4\% of the internal flux). Relatively similar conclusions were reached by Bouchy et al. (2011) for CoRoT-15b even though it was a less massive brown dwarf receiving a more intense irradiation.

Currently, the most widely accepted explanation for the radius anomaly of low-mass stars is through inhibition of convection by strong magnetic fields \citep[][]{mullanmacdonald2001}. This explanation has recently been shown to be satisfactory for three well-characterized systems \citep[][]{macdonaldmullan2014}, although the internal magnetic field needed might be too strong compared with analytical arguments \citep[][]{feidenchaboyer2014}. However, this scenario usually requires a relatively fast rotation rate to maintain a dynamo - fast rotation rate that is enforced by tidal synchronization in short-period binaries.

For \189 B, the timescale for tidal de-spinning is expected to be longer than $10-100$ Gyr, so that synchronous rotation (that would be too slow anyway) is most likely never reached.
But this does not necessarily imply a slow rotation. Indeed, the relatively low effective and equilibrium temperatures of \189 B imply a very poorly ionized atmosphere \citep{mohanty2002}, which causes a decoupling of the flow and magnetic field, yielding a reduced activity. This, in turn, should result in reduced winds and inefficient magnetic braking. If there is no other significant angular-momentum transport mechanism, it is then possible that \189 B retained its initial angular momentum. Because of the continuous gravitational contraction, this could result in rotation periods as short as a few hours \citep{herbst2007}. Magnetically inhibited convection is therefore still a possible explanation. In addition, \citet{chabrier2007} suggested that spot coverage could substantially increase the radius of magnetically active low-mass stars, especially in a binary system. However, for the reasons discussed above, the relatively low effective temperature is expected to yield a reduced activity and thus a smaller spot coverage. 

Since a rapid rotation cannot be excluded, one has to take into account the deformation and the net centrifugal expansion of the object \citep{leconte2011}. For a rotation period of 1 hr (a third of the breakup velocity), and using Eq.\,(33) of \citet{leconte2011} along with their tabulated values of the structural properties of substellar objects and very low-mass stars, we derive an increase in the mean radius of 2.5\%. Although a significant effect considering the uncertainties, this effect is not sufficient by itself.

Finally, a possible explanation would be that the \189 system is younger than expected ($\leq$1 Gyr). Although this age is not discarded by the combined analysis of the SOPHIE spectra and the SED measurements, the $v \sin i$ estimated from the spectrum points to a slowly rotating primary star, which would indicate instead a more advanced age.

In conclusion, because of its very peculiar location at the very bottom of the main sequence and its relatively wide orbit, \189 B is among the most challenging objets to explain and therefore provides a strong constraint on future structure models.

 \subsection{Additional companions}
The detection of additional companions in these systems can provide insights on their formation and evolution history \citep[e.g.,][]{luhman2012, burgasser2007}. 

\begin{figure*}
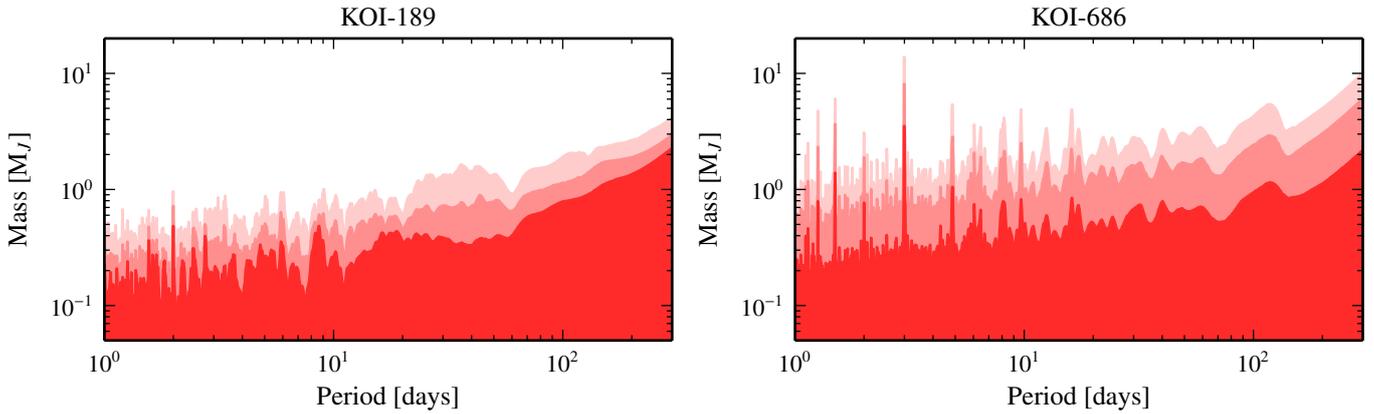

\input{KOI-189_limits2ndplanet.pgf}
\input{KOI-686_limits2ndplanet.pgf}
\caption{Upper limits to for the mass of a putative second companion in the system as a function of orbital period. The shaded regions correspond to the 68.3\% (dark), 95.5\%, and 99.7\% (light) confidence level. \label{fig.uppermasslim}}
\end{figure*}

To set upper limits to the mass of these putative companions, we considered a random set of 1000 posterior samples obtained with the MCMC algorithm and computed the residuals for each set of parameters in the set. The best-fit (in the least-squares sense) amplitude and corresponding mass were obtained for each sample, and for a series of periods from 1 to 300 days. From the mass distributions thus obtained, we computed the 68.3th, 95.3th and 99.7th percentiles (Fig.~\ref{fig.uppermasslim}). For \189, companions more massive that 1 \MJ\ were discarded up to orbital periods of around 20 - 60 days, which corresponds to half the time span of the observations. For longer periods, the upper mass limit increases linearly with the period. In the \686\ system, the upper limit is slightly less stringent, possible because there are fewer RV points. Still, no companions with masses above a few Jupiter masses are allowed for periods shorter than around 60 days (again, half the time span of RV observations). Additional data and long-term follow-up of these systems will make these limits more stringent and enrich our understanding of very low-mass stars.

\begin{acknowledgements}

We thank the staff of Haute-Provence Observatory for their 
support at the 1.93-m telescope and on \sophie.
We thank the ``Programme National de Plan\'etologie'' (PNP) of CNRS/INSU, 
the Swiss National Science Foundation, 
and the French National Research Agency (ANR-08-JCJC-0102-01 and ANR-NT05-4-44463) 
for their support with our planet-search programs.

This publication makes use of data products from the Wide-field Infrared Survey Explorer, which is a joint project of the University of California, Los Angeles, and the Jet Propulsion Laboratory/California Institute of Technology, funded by the National Aeronautics and Space Administration.

This research was made possible through the use of the AAVSO Photometric All-Sky Survey (APASS), funded by the Robert Martin Ayers Sciences Fund.

The team at LAM acknowledges support by CNES grants 98761 (SCCB), 426808 (CD), and 251091 (JMA). AS acknowledge the support from the Euro- pean Research Council/European Community under the FP7 through Starting Grant agreement number 239953. RFD carried out part of this work thanks to the support by CNES via its postodoctoral fellowship program. 

AS is supported by the European Union under a Marie Curie Intra-European Fellowship for Career Development with reference FP7-PEOPLE-2013-IEF, number 627202.

ASB and RFD acknowledge funding from the European Union Seventh Framework Programme (FP7/2007-2013) under Grant agreement n. 313014 (ETAEARTH)

The authors gratefully acknowledge the anonymous referee for her/his comments and suggestions for improving the article.
\end{acknowledgements}

\bibliographystyle{aa}

\end{document}